\def\cm2{cm$^{-2}$}

\def\c2{C~{\sc ii}}
\def\c4{C~{\sc iv}}
\def\fe2{Fe~{\sc ii}}
\def\fe3{Fe~{\sc iii}}
\def\mg1{Mg~{\sc i}}
\def\mg2{Mg~{\sc ii}}
\def\si2{Si~{\sc ii}}
\def\si4{Si~{\sc iv}}
\def\al2{Al~{\sc ii}}
\def\al3{Al~{\sc iii}}
\def\o1{O~{\sc i}}
\def\n1{N~{\sc i}}
\def\h1{H~{\sc i}}

\def\approxlt{\mathrel{\spose{\lower 3pt\hbox{$\sim$}}
        \raise 2.0pt\hbox{$<$}}}
\def\approxgt{\mathrel{\spose{\lower 3pt\hbox{$\sim$}}
        \raise 2.0pt\hbox{$>$}}}

\def\lea{\mathrel{<\kern-1.0em\lower0.9ex\hbox{$\sim$}}}
\def\gea{\mathrel{>\kern-1.0em\lower0.9ex\hbox{$\sim$}}}

\documentclass[article]{gwp80}
\usepackage{graphicx}
\usepackage{amssymb}
\tabletypesize{\normalsize}  

\shortauthors{Sarajedini}
\shorttitle{RR Lyraes in M31 and M33}

\begin{document}
\large    
\pagenumbering{arabic}
\setcounter{page}{1}

\title{RR Lyrae Variables in M31 and M33}

%
%
\author{{\noindent Ata Sarajedini} \\
\\
{\it Department of Astronomy, University of Florida, 211 Bryant Space
Science Center, Gainesville, FL 32611, USA} 
}
%
%
\email{ata@astro.ufl.edu}


\begin{abstract}
The properties of RR Lyrae variables make them excellent probes of
the formation and evolution of a stellar population. The mere presence
of such stars necessitates an age greater than $\sim$10 Gyr while
their periods and amplitudes can be used to estimate the metal
abundance of the cluster or galaxy in which they reside.  These
and other features of RR Lyraes have been used to study the
properties of M31 and M33. Though these studies
are generally in their infancy, we have established that M31 and M33 do
indeed harbor RR Lyraes in their halos and probably also in their disks
suggesting that these two components formed early in the history
of M31 and M33. The mean metallicities of the halo RR Lyraes in
these galaxies are consistent with those of other halo stellar population 
tracers such as the dwarf spheroidal satellites of M31 and the halo globular 
clusters in M33. Little is known about the spatial distribution of the RR Lyraes,
especially in M33.
This will require wide-field time-series studies with sufficient
photometric depth to allow both the identification of RR Lyraes and
robust period determination.
\end{abstract}

\section{Introduction}

The class of pulsating stars known as RR Lyrae variables are located
at the intersection of the instability strip and the horizontal branch in the
Hertzsprung Russell Diagram. Their utility has been widely documented
in the literature. As such, they can be referred to as the ``swiss army knife" 
of astronomy as graphically illustrated in Fig. 1. 

Because of their low masses ($\sim$0.7 M$_\odot$, Smith 1995), 
the mere presence of RR Lyrae stars in a stellar population 
suggests an old age ($\gea$10 Gyr) 
for the system. As such, one does not need to obtain
deep photometry beyond the old main sequence turnoff in order
to establish the presence of an old population. Generally
speaking, {\it identifying} RR Lyrae variables does not require a
substantial investment of telescope time. 

\begin{figure*}
\centering
\includegraphics[width=10cm,angle=90]{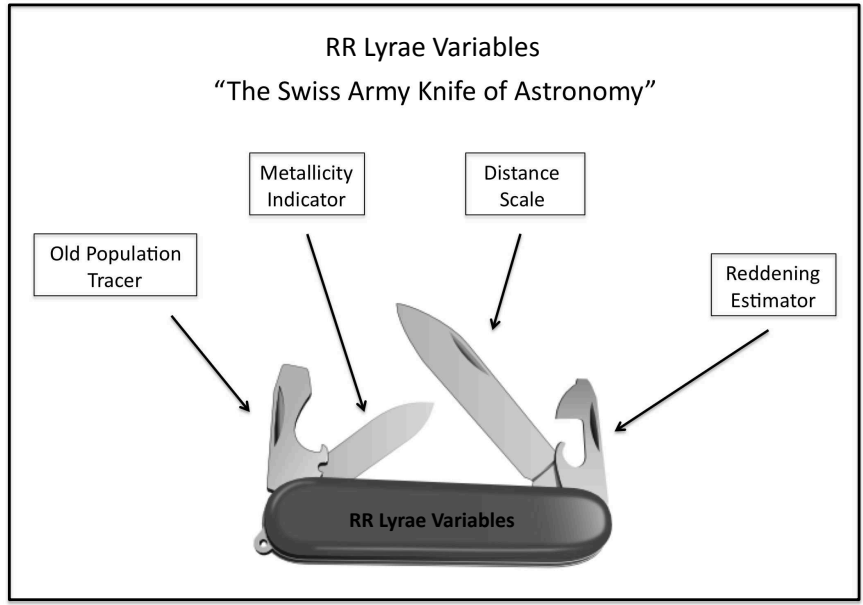}
\vskip0pt
\caption{Graphic illustrating the multiple uses of RR Lyrae variables in
astronomy. They are well known as distance indicators, and, in addition, their 
mere presence necessitates the existence of a stellar population
with an age greater than $\sim$10 Gyr.
Their periods and amplitudes are useful for measuring the metal
abundance of the star cluster or galaxy in which they reside.  Furthermore, their
minimum light colors can be used to measure the line-of-sight reddening.}
\end{figure*}

There are three principal types of RR Lyrae variables; those pulsating in the
fundamental mode exhibit sawtooth-like light curves and are referred to as
ab-type or RR0 variables. The first overtone pulsators generally show sine-curve
shaped light curves, have smaller periods and typically lower amplitudes than the
ab-types, and are referred to as c-type or RR1 variables. Lastly, RR Lyraes that pulsate
in both the fundamental and first overtone modes (i.e. double mode pulsators) 
carry the d-type moniker.

The periods of the ab-type RR Lyrae stars (P$_{ab}$) are related to their
metallicities. Using data on field RR Lyraes from
Layden (2005, private communication), Sarajedini et al. (2006)
found 
\begin{eqnarray}
[Fe/H] = -3.43 - 7.82~Log~P_{ab}.
\end{eqnarray}
\noindent The dispersion
in this relation (rms = 0.45 dex) is significant making the determination of 
individual stellar metallicities unreliable, but the relation is useful for 
estimating the mean abundance of a population of RR Lyraes. There 
is a more precise relation given by Alcock et al. (2000) that requires
knowledge of the period {\it and} amplitude [A(V)]. They found
\begin{eqnarray}
[Fe/H] = -8.85[Log~P_{ab}-0.15A(V)] - 2.60.
\end{eqnarray}
\noindent With this relation, the error per star
is reduced to $\sim$0.31 dex and the precision of the resulting abundance
distribution is narrower (Sarajedini et al. 2009). 

Once the metallicities of the RR Lyraes are determined, their
absolute magnitudes can be calculated. The published equations typically
take the form of a linear relation between [Fe/H] and M$_V$(RR). 
A number of different slopes and zero-points have been derived 
for this equation, but there seems to be convergence on slope
values of $\sim$0.20 and zero-points of $\sim$0.90 
(Chaboyer 1999, Gratton et al. 2003, 2004; Dotter et al. 2010).

The work of Bono et al. (2007) has combined the process of metallicity and
distance determination together into one coherent technique. They describe a
process that uses the periods of the ab-type RR Lyraes along with their amplitudes to
calculate absolute magnitudes, which are then input into a relation between
M$_V$(RR) and [Fe/H] in order to determine the metal abundance. In applying this
method to their observations of M31 RR Lyraes,
Jeffrey et al. (2011) claim that the Bono et al. (2007) method for
deriving metallicities from period and amplitude data is superior to those
of Alcock et al. (2000) and Sarajedini et al. (2006).

Moving on to the reddening of the RR Lyraes,
the minimum light colors of ab-type RR Lyraes are largely 
independent of their other properties as shown by Guldenschuh et al.
(2005) and Kunder et al. (2010). 
This is based on a concept originally developed by Sturch (1966).
As a result, if the minimum light colors are well-determined, they can
be compared with $(V-I)_{o,min} = 0.58 \pm 0.02$ and
$(V-R)_{o,min} = 0.28 \pm 0.02$\footnote{Depending on how the
minimum light color is measured, this value could also be 
$(V-R)_{o,min} = 0.27 \pm 0.02$. See Kunder et al. (2010) for 
details.} in order to 
measure the line-of-sight reddening for each star.

Thus far, we have presented evidence for how RR Lyrae variables 
can be powerful probes of
the systems in which they reside - star clusters or among the field
populations of galaxies. It is for this reason that studying them in
the Local Group spiral galaxies M31 and M33 provides valuable
insights into the properties of these systems. Ultimately, we would like to
know how large spiral galaxies like M31 and `dwarf spirals' like M33 fit 
into the process of galaxy formation 
in a Universe dominated by cold dark matter (CDM) with a 
cosmological constant $\Lambda$ (Navarro, Frenk,  \& White 1997).
Comprehensive knowledge of the most ancient stars in these systems
will shed light on this question. In the remainder of this contribution, we will
describe how RR Lyrae stars have been used for this purpose.

%

\section{RR Lyraes in M31}

The field RR Lyrae population of M31 was first systematically studied by Pritchet \&
van den Bergh (1987).  They used the Canada-France-Hawaii 3.6m telescope
to observe a field at a distance of 9 kpc from the center of M31 along the minor
axis partially overlapping the field observed by the seminal work of Mould \& Kristian (1986). 
Pritchet \& van den Bergh (1987) identified 30 RR Lyrae candidates and were able to 
estimate periods for 28 of them. These ab-type variables have a mean period of 
$\langle$$P_{ab}$$\rangle$=0.548 days. The
photometric errors in their data prevented them from identifying the lower-amplitude
c-type RR Lyraes. 

Dolphin et al. (2004) observed the same field as Pritchet
\& van den Bergh (1987) using the WIYN 3.5m on Kitt Peak. They found
24 RR Lyrae stars with a completeness fraction of 24\%, suggesting that their
$\sim$100 square arcmin field could contain $\sim$100 RR Lyraes resulting in a
density of about one RR Lyrae per square arcmin. This is much less than the value of
$\sim$17 per square arcmin found by Pritchet \& van den Bergh (1987). They also
noted  that the mean metallicity of the M31 RR Lyraes seemed to
be significantly lower than that of the M31 halo. The work of Durrell et al.
(2001) had reported a peak value of $[M/H]$$\sim$--0.8 for the M31 halo, and
Dolphin et al. (2004) were not able to reconcile this abundance value with the distance
implied by the mean magnitude of their RR Lyrae sample. 

The most definitive work on the RR Lyraes of M31 was published by Brown et al.
(2004) and made use of $\sim$84 hours of imaging time 
(250 exposures over 41 days) with the Wide Field Channel of
the Advanced Camera for Surveys (ACS/WFC) onboard HST. Their 
field was located along the minor axis of M31 approximately 11 kpc from
its center. Their analysis revealed a complete sample of RR Lyrae stars
consisting of 29 ab-type variables and 25 c-type. Using equation (2) above,
the periods and amplitudes of these
stars suggest a mean metallicity of [Fe/H]$\sim$-1.7 for the old
population in the Andromeda halo. This is qualitatively consistent with
the assertions of Dolphin et al. (2004) regarding the metal abundance of
the M31 halo - that it is lower than the value suggested by the work of
Durrell et al. (2001). As we discuss below, more recent work has shown that the 
M31 bulge dominates the spheroid inside $\sim$30 kpc while the halo dominates 
from $\sim$30 to $\sim$165 kpc (Guhathakurta et al. 2005; 
Irwin et al. 2005) and has a metallicity that is comparable to that of the Milky 
Way halo (Kalirai et al. 2006; Koch et al. 2008). 

\begin{figure}
\includegraphics[width=14cm]{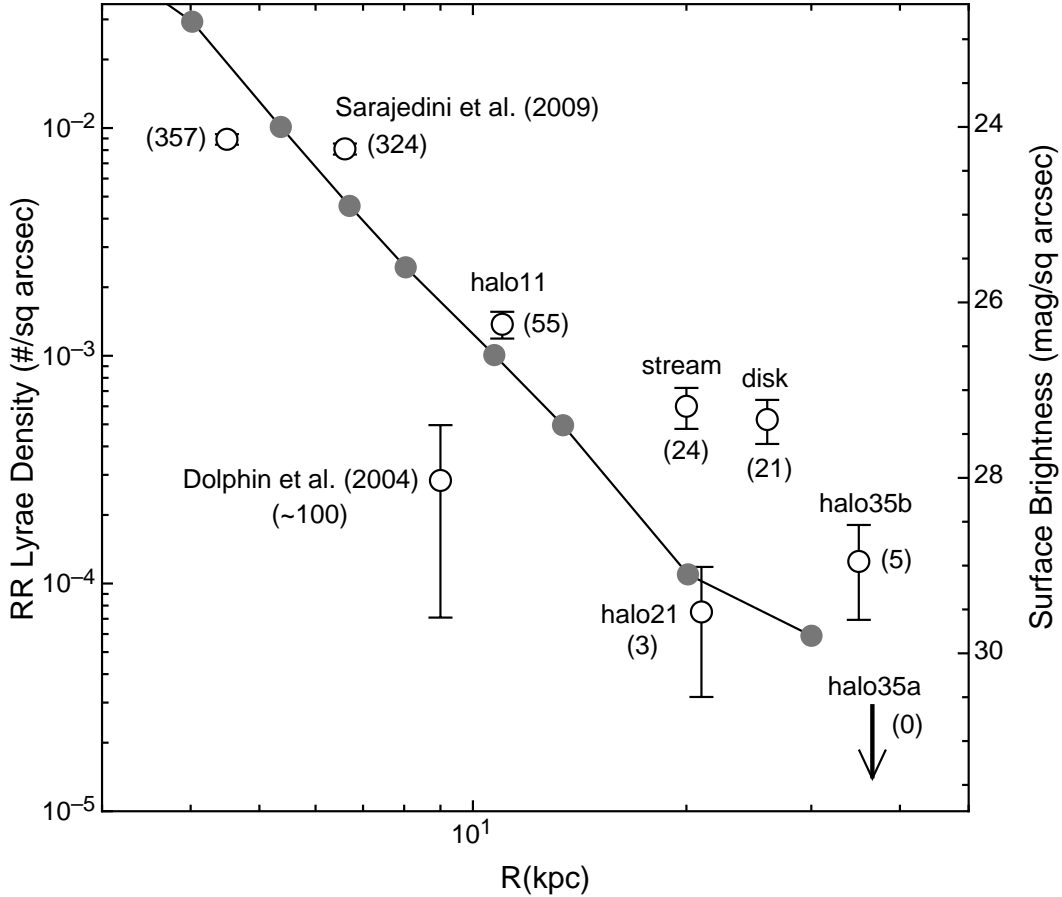}
\centering
\caption{The open circles represent the surface density of RR Lyrae variables
identified in M31 with each point labeled based on its source. Those designated
halo11, stream, disk, halo21, and halo35b are taken from the work of Jeffrey et al.
(2011).  Note that the Dolphin et al. (2004) value, which is based on WIYN 3.5m
telescope MiniMosaic observations, represents an extrapolation
based on their estimate of the incompleteness of their sample. The remaining
points are all taken from HST/ACS studies. The values in parenthesis indicate
the raw number of RR Lyraes in each field. The solid points connected by line
segments show the surface brightness profile of M31 taken from Fig. 4b of Pritchet
\& van den Bergh (1994) scaled to match the average of the inner two RR Lyrae
density points. Neglecting the point based on the work of Dolphin et al. (2004), it
would appear that the stream and disk fields exhibit an excess of RR Lyraes above
that of the M31 halo.} 
\end{figure}

The paper by Jeffrey et al. (2011), which is the most recent contribution in this
area, is a continuation of the Brown et al. (2004) 
work and presents high quality light curves for RR Lyraes in 5 additional fields around
M31. These include a halo field at 21 kpc, and two halo fields at 35 kpc. In addition, there
is a field coincident with one of the streams in the vicinity of M31 and one that covers
a disk region about 26 kpc from the center along the major axis of the galaxy. 
Figure 2 presents the radial density profile of M31 RR Lyraes from the studies
mentioned thus far. These are compared with the surface brightness profile illustrated
in Fig. 4b of Pritchet \& van den Bergh (1994). The latter has been normalized to
match the average of the inner two RR Lyrae points. Neglecting the point based 
on the work of Dolphin et al. (2004), which is an extrapolation based on a
completeness correction applied to their sample, it
would appear that the stream and disk fields in M31 exhibit an excess of RR Lyraes 
above that of the M31 halo. The surface brightness profile of the halo predicts
the presence of $\sim$4 halo RR Lyraes in the stream field and $\sim$3 such stars
in the disk field leaving $\sim$20 and $\sim$18 RR Lyraes in the stream and disk,
respectively. If this finding withstands further scrutiny, it suggests that the 
oldest stars in the M31 disk, halo, and stream all have very similar ages and
metallicities. 
Jeffrey et al. (2011) also performed a comparison of the three methods described
in Sec. 1 to estimate the mean metallicity of an RR Lyrae population. They conclude
that the three methods are roughly consistent with each other in terms of the
derived mean metallicity of the RR Lyrae population.

Sandwiched between the Brown et al. (2004) paper and its successor, the work of 
Jeffery et al. 2011, is the HST/ACS study of Sarajedini et al. (2009), which is also
focused on the RR Lyrae population of M31. A total of 681 RR Lyraes (555 ab-type 
and 126 c-type) were identified in two fields located at
$\sim$4 kpc and $\sim$6 kpc from the center of M31. A mean metal abundance
of [Fe/H]$\sim$-1.5 was determined using equation (2) above. 
In addition, as the Bailey diagrams shown in Fig. 10 of Sarajedini et al. (2009) and 
Fig. 12 of Jeffrey et al. (2011) illustrate, the RR Lyrae stars in M31 are largely 
consistent with the those in Oosterhoff (1939) type I Galactic globular clusters. 

\begin{figure}
\includegraphics[width=14cm]{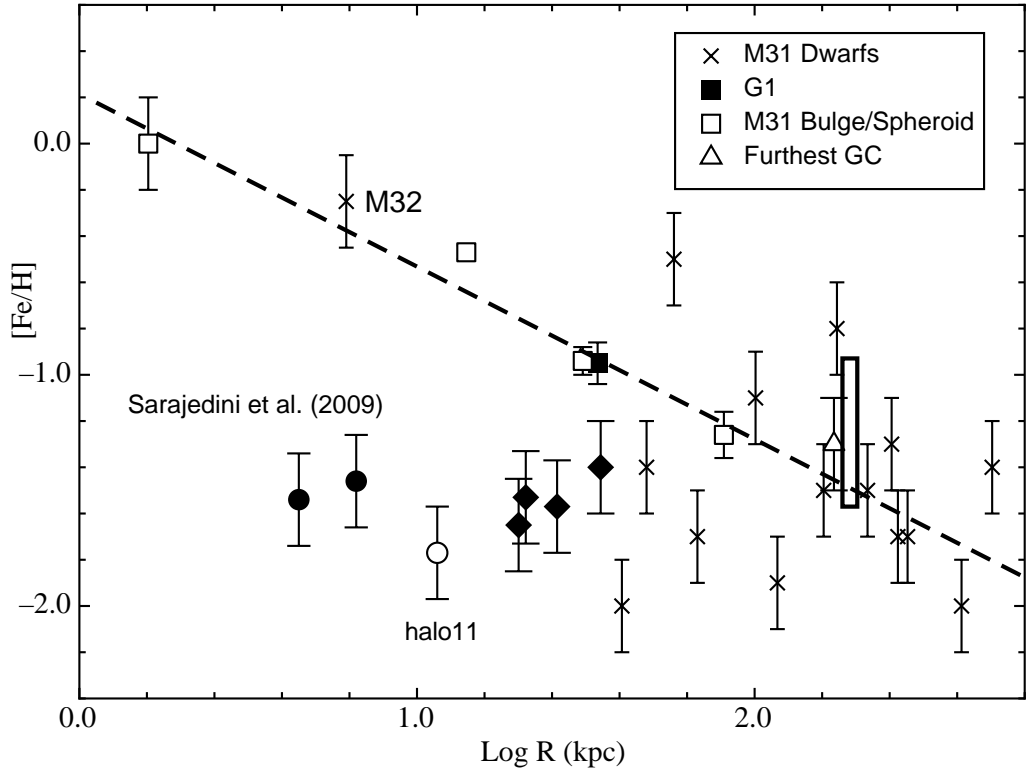}
\centering
\caption{A plot of the variation of metal abundance with projected distance from the 
center of M31. The
inner most open square represents the bulge abundance measured by Sarajedini \&
Jablonka (2006). The remaining open squares are the bulge/halo points from the work of
Kalirai et al. (2006). The dashed line is the least squares fit to these data with a slope
of --0.75 $\pm$ 0.11. The crosses
represent the dwarf galaxies surrounding M31 from the work of Grebel et al.
(2003) and Koch \& Grebel (2006) whereas the abundance of M32 is taken from
Grillmair et al. (1996). The filled square is the well-known massive globular cluster
G1 studied by Meylan et al. (2001). The open triangle is the furthest known globular cluster
in M31 discovered by Martin et al. (2006). The points from Sarajedini et al. (2009, filled
circles) and Brown et al. (2004, open circle) are also plotted. The diamonds are the
fields observed by Jeffrey et al. (2011) and designated halo21, stream, disk, and halo35b.
Unless otherwise noted, an error of 0.2 dex is adopted for each point derived from
the RR Lyraes.
For completeness, the boxed region shows the location of the halo globular clusters in M33
from the work of Sarajedini et al. (2000). All of these points have been scaled to an
M31 distance of $(m-M)_0$ = 24.43.}
\end{figure}

Seeking to place the mean metallicity of its RR Lyraes within the broader context of
the radial metallicity gradient in M31 and its environs, Sarajedini et al. (2009)
constructed the plot shown in Fig. 3. The inner-most point in Fig. 3 is the bulge 
metallicity from the work of
Sarajedini \& Jablonka (2005), while the remaining open squares are the bulge/halo
points from the work of Kalirai et al. (2006) as shown in their Table 3. The dashed
line is the least squares fit to the open squares. The other points represent
the dwarf spheroidal companions to M31 (crosses, Grebel et al. 2003; 
Koch \& Grebel 2006), the globular cluster G1 (filled square, Meylan et al. 2001),
and the furthest globular cluster in M31 (open triangle, Martin et al. 2006). Note
that we have adopted the mean metallicity of M32 from the work of Grillmair
et al. (1996). The metallicities for the RR Lyraes in the two fields observed
by Sarajedini et al. (2009) are shown by the filled circles while the open circle 
represents the RR Lyraes in the halo11 field of Brown et al. (2004) field. The
filled diamonds are the halo21, stream, disk, and halo35b fields in order of
increasing galactocentric distance from Jeffery et al. (2011). For completeness, the
elongated rectangle represents the locations of the halo globular clusters
in M33 from Sarajedini et al. (2000). All of these values are based on a 
distance of $(m-M)_o = 24.43$ (770 kpc) for M31. In
cases where an error in the metallicity is not available, we have adopted a value
of $\pm$0.2 dex.

We see in Fig. 3 a clear representation of the notion that the halo population
in M31 does not begin to dominate until a galactocentric distance of $\sim$30 kpc,
as suggested by a number of authors (Guhathakurta et al. 2005; Irwin et al. 2005;
Kalirai et al. 2006; Koch et al. 2007). At this location, we see
a transition region between the globular cluster G1 which is consistent with the 
inner-spheroid metallicity gradient (dashed line) and the dwarf spheroidal
galaxies which show no relation between abundance and galactocentric
distance. 
In this sense, it would appear that the RR Lyrae populations follow the trend outlined 
by the stellar populations
outside of $\sim$30 kpc. This indicates that all of the RR Lyraes, with the
likely exception of those in the disk and stream fields,
are probably members of the M31 halo rather than its bulge suggesting that
the halo can be studied as close as 4 kpc from the center of M31 by
focusing on the RR Lyraes.

\section{RR Lyraes in M33}

The history of RR Lyrae studies in M33 is relatively short as compared with
M31. The earliest
study is that of Pritchet (1988), who presented preliminary results for
a handful of such stars. No data or light curves were shown, but
Pritchet (1988) did estimate a distance of 
(m--M)$_{0}$ = 24.45 $\pm$ 0.2 for M33 based on the RR Lyrae
variables. This value is somewhat smaller than the average of several
different determinations from Galleti et al. (2004) of 
(m--M)$_{0}$ = 24.69 $\pm$ 0.11. 

The first study to unequivocally identify and characterize RR Lyraes in
M33 was that of Sarajedini et al. (2006). They used time-series
observations of two fields in M33 taken with ACS/WFC onboard HST. 
The observations consisted of 8 epochs 
in the F606W ($\sim$V) filter and 16 epochs in the F814W ($\sim$I)
filter. The data were analyzed using the light curve-template-fitting software 
developed by Andrew Layden and described in Layden \& Sarajedini
(2000, see also Mancone \& Sarajedini 2008). Based on these data, 
64 ab-type RR Lyraes were identified.
However, very few c-type variables were uncovered because of their
generally lower amplitude.

The period distribution of the ab-type variables showed two peaks -
one at longer periods which resembles the metal-poor 
RR Lyraes in M3 and M31 (Brown et al. 2004) and one at shorter 
periods which could be from metal-rich RR Lyraes in M33's disk. 
The presence of the latter population is somewhat uncertain given
the recent work of Pritzl et al. (private communication). Sarajedini
et al. (2006) found the mean
metallicity of the metal-poor RR Lyraes to be consistent
with that of halo globular clusters in M33 which have
$\langle$[Fe/H]$\rangle$ = --1.27 $\pm$ 0.11 (Sarajedini et al. 2000).

Figure 12 of Sarajedini et al. (2006) shows that the M33 RR Lyraes are
in their expected location in the color-magnitude diagram (CMD);
in addition, their colors exhibit a dispersion that is consistent
with being significantly affected by differential reddening. This suggests
that some of the RR Lyraes are on the near side of M33 while
others are in the disk and/or on the far side of the galaxy. One way
to further investigate this possibility is to examine the distribution
of RR Lyrae reddenings using the intrinsic minimum light color
of the ab-types as mentioned in Sec. 1. This analysis was performed
by Sarajedini et al. (2006) and shows that the RR Lyraes span
the range from E(V--I)$<$0.1 up to E(V--I)$\sim$0.7. Given that the
line-of-sight reddening to M33 is E(V--I)$\sim$0.1, this suggests
that RR Lyraes exist in the disk of M33 and in its halo (on the near
and far side). We return to this point later as we discuss the most
recent results on RR Lyraes in M33.

The primary result from the work of Sarajedini et al. (2006) was
that M33 does indeed contain RR Lyrae variables in its halo. This
suggests that the halo of M33 contains some fraction of stars with
ages older than $\sim$10 Gyr. In this regard, the halos of M33,
M31, and the Milky Way are similar. It seems that they started forming
stars at about the same time. 

\begin{figure}
\centering
\includegraphics[width=18cm]{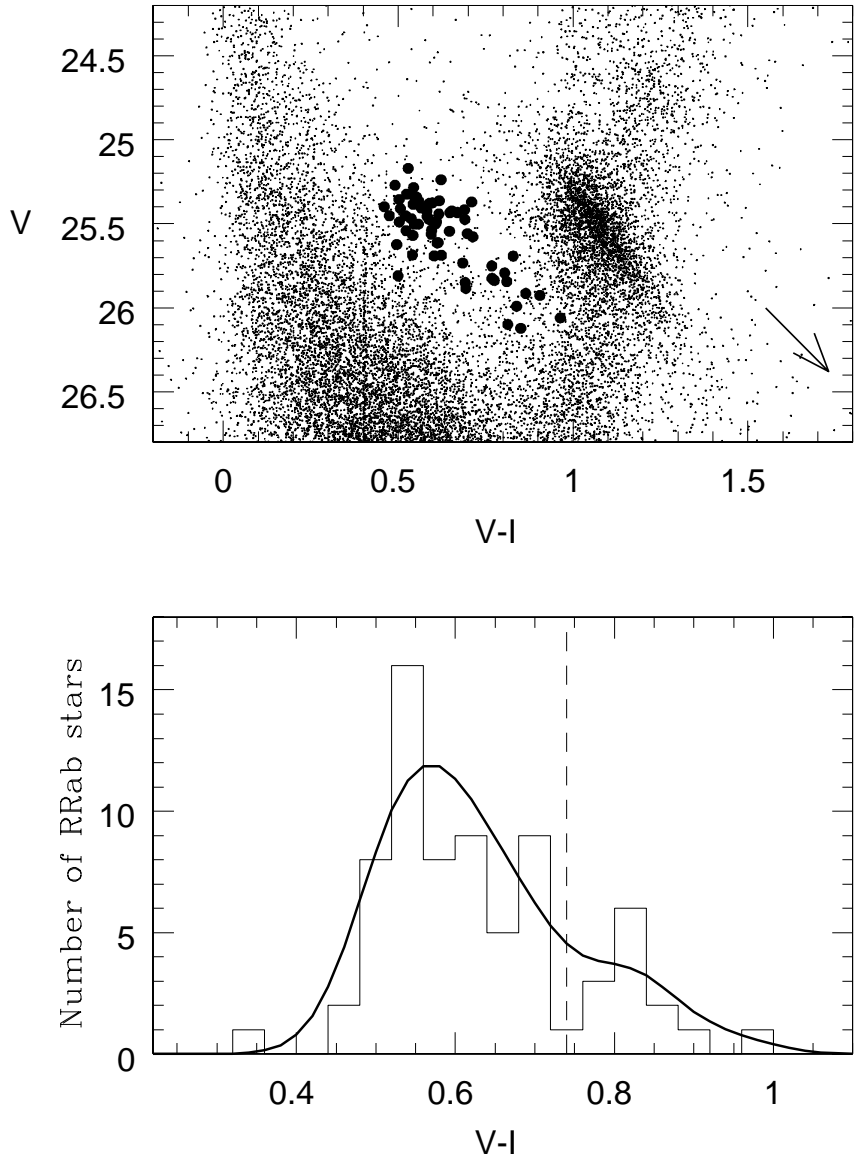}
\caption[]{The upper panel shows the color magnitude diagram of
the DISK2 field from the work of Yang et al. (2010) along with the
RR Lyraes identified in this field (filled circles). The brightest two or three stars
are likely to be anomalous cepheids. The arrow shows the reddening
vector in the CMD. The lower panel illustrates the color histogram
for the RR Lyrae variables, which appears to show a bimodal 
distribution with a primary peak at (V--I)$\sim$0.5 and a secondary
one at (V--I)$\sim$0.8 with a minimum around (V--I)=0.74 (dashed
line). } 
\end{figure}

\begin{figure}
\centering
\includegraphics[width=18cm]{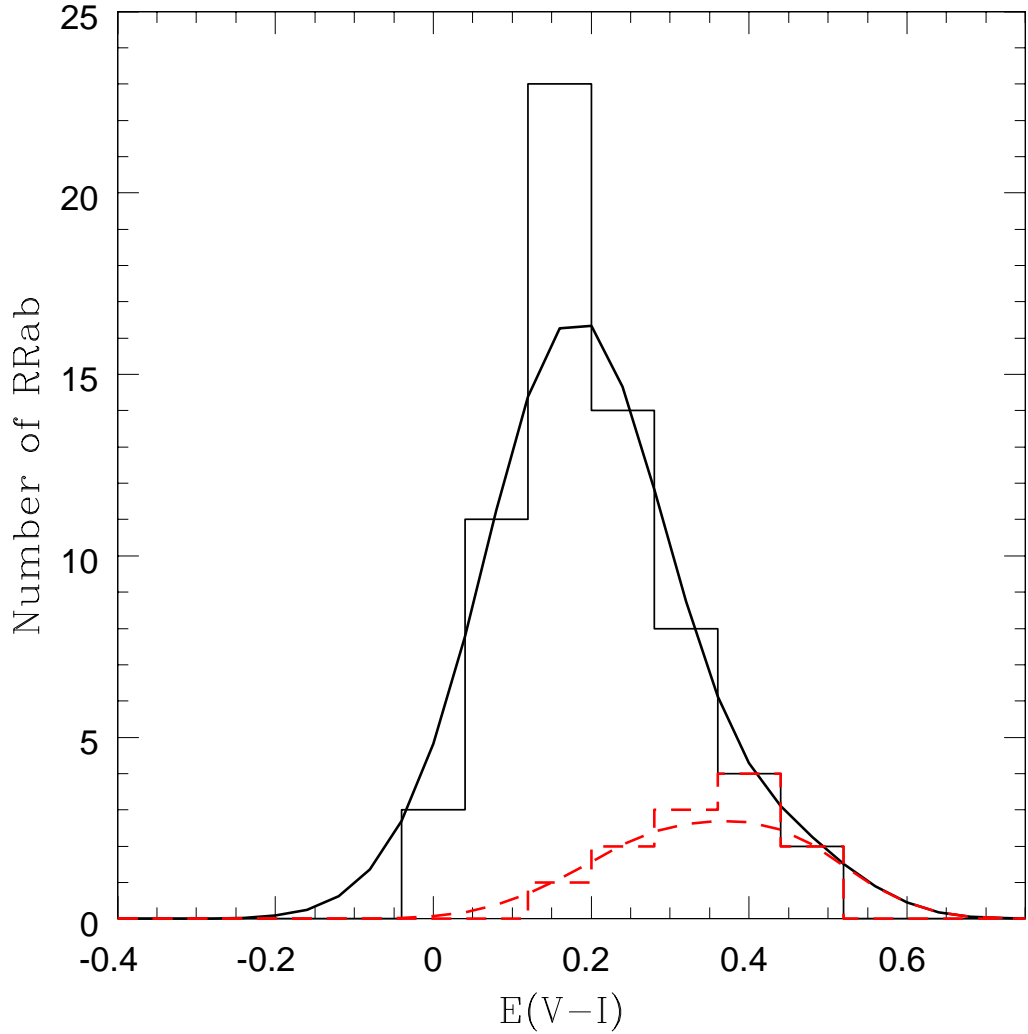}
\caption[]{The solid line is the distribution of reddenings for ab-type RR 
Lyraes identified in M33 from the study of Yang et al. (2010). The 
dashed line is the reddening distribution of RR Lyraes with 
(V--I)$>$0.74 (dashed line in Fig. 4) showing that these stars
are red because of reddening in the disk of M33.} 
\end{figure}

\begin{figure}
\includegraphics[width=20cm]{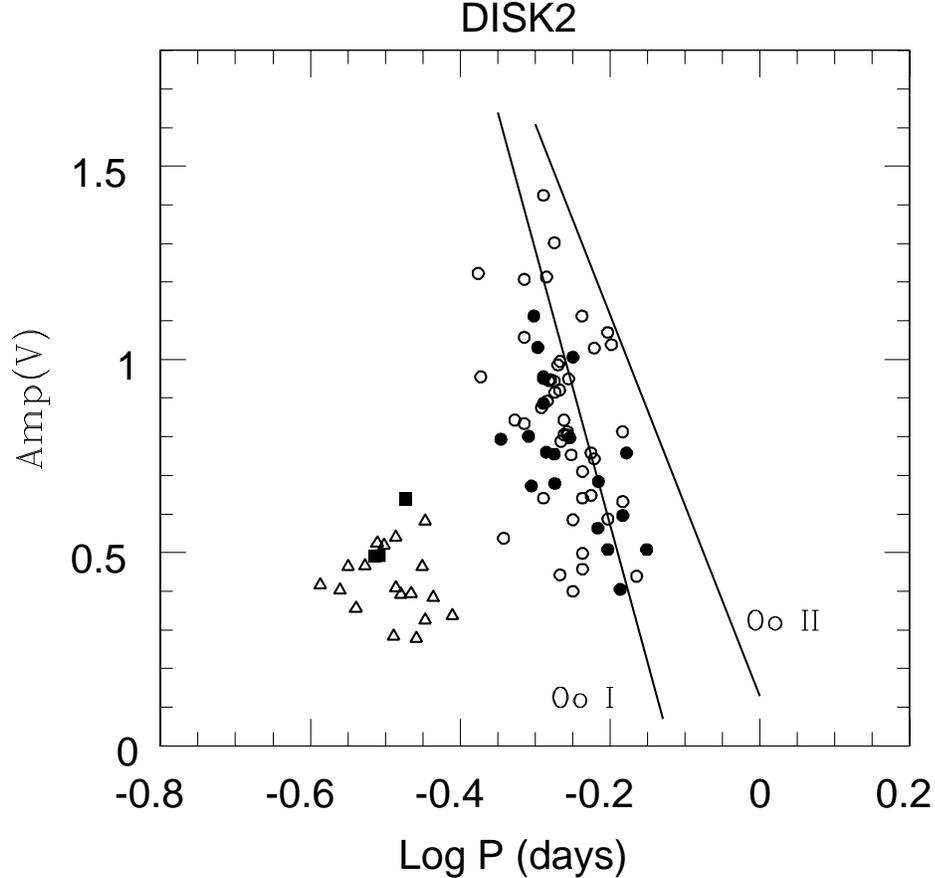}
\caption[]{This shows the Bailey Diagram for all RR Lyraes in
the M33 DISK2 field. The open circles are the ab-type variables while
the open triangles are the c-types. The few candidate d-type RR Lyraes
are identified with filled squares. The solid lines are the locations of the
Oosterhoff type I and Oosterhoff type II Galactic globular clusters from
the work of Clement \& Rowe (2000). The filled circles are the ab-type
RR Lyraes with (V--I)$>$0.74 from Fig. 4; these are likely to be located
on the far side of M33's disk. See text for discussion.} 
\end{figure}

\begin{figure}
\includegraphics[width=17cm]{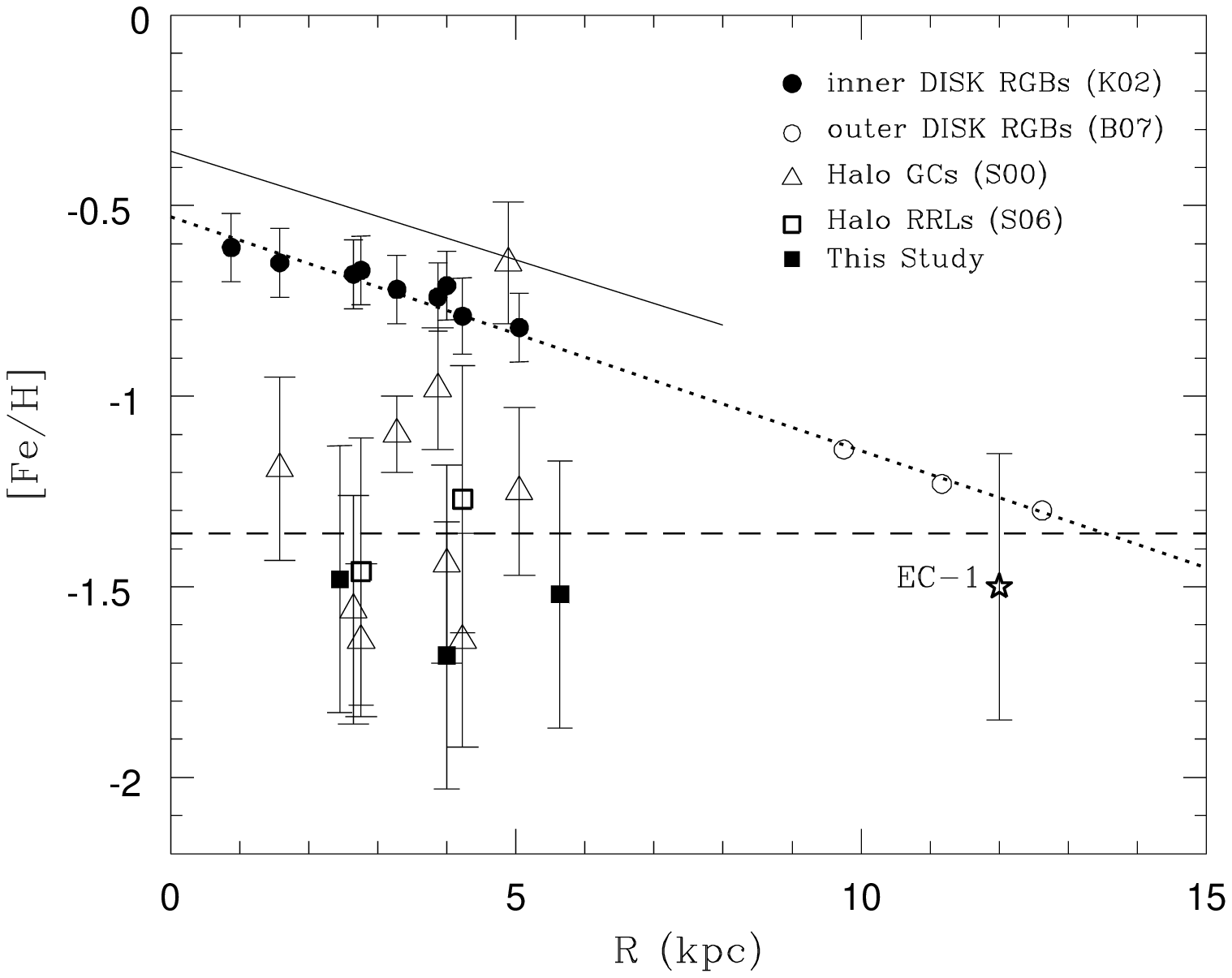}
\caption[]{The metallicity of the M33 stellar populations as a function of deprojected 
radial distance from the galactic center taken from Yang et al. (2010). 
The dotted line is a least-squares fit to the 
radial trend of the M33 disk red giant branch stars from Kim et al. (2002, filled circles)
and also happens to coincide with the outer disk points from Barker et al. (2007,
open circles). The solid 
line shows the metallicity trend of the M33 planetary nebulae (PNe) from Magrini et al. 
(2010). The dashed line represents the mean metallicity ([Fe/H] = Ð1.36) of the 
M33 halo stellar populations in this plot (globular cluster - open triangles,
RR Lyraes - squares). The open star represents EC-1, which 
is the furthest known star cluster from the center of M33 (Stonkute et al. 2008).
Note that ``This Study" refers to the work of Yang et al. (2010) from which
this figure was taken.} 
\end{figure}

Building upon the work of Sarajedini et al. (2006), Yang et al. (2010)
present an analysis of new HST/ACS/WFC imaging that is part of 
program GO-10190. The primary aim of this program is to study
the star formation of the disk of M33 (Williams et al. 2009). 
As such, fields were obtained
at four different disk locations roughly equally spaced along the 
major axis of M33. Figure 1
of San Roman et al. (2009) shows the locations of the disk fields. Here
we focus on the properties of the RR Lyraes in the second closest field
to the center of the M33, which we designate DISK2. 

The observations are composed of 16 epochs in the F606W filter
and 22 in the F814W filter spanning a time window of $\sim$3 days.
Template-fitting analysis suggests the presence of 86 RR Lyraes
in this field - 65 ab-type, 18 c-type, and 3 d-type variables. The upper panel of
Figure 4 shows the CMD of DISK2 along with the locations of the
RR Lyraes. The lower panel of Fig. 4 displays the distribution in color
of the ab-type RR Lyraes revealing the presence of two peaks - 
a primary peak at (V--I)$\sim$0.5 and a secondary
one at (V--I)$\sim$0.8. We would like to know if this bimodality is
due to reddening internal to M33. As such, we determine the
reddening of each RR Lyrae using Sturch's method as described in Sec. 1
and then examine the distribution of reddenings to see if the fainter/redder
RR Lyraes, those with (V--I)$>$0.74, do indeed suffer from higher reddening.
Figure 5 illustrates this effect. The
solid histogram represents all ab-type RR Lyraes while the dotted
histogram shows only those with (V--I)$>$0.74 (dashed line in the bottom
panel of Fig. 4). Our hypothesis seems to be correct - that the fainter/redder 
RR Lyraes are being affected by extinction internal to M33. This
suggests that the variables near (V--I)$\sim$0.5 are on the near
side of M33 while those with (V--I)$\sim$0.8 are in the disk or on the far side.

We now seek to compare the properties of the low reddening and high reddening
RR Lyraes. In particular, how do they compare in the Bailey Diagram?
This is shown in Fig. 6 where we plot the periods
and amplitudes of the DISK2 RR Lyraes (open circles - ab-types,
open triangles - c-types; filled squares - d-types; filled circles - higher
reddening ab-types). These are compared
with the Oosterhoff I and II loci from Clement \& Rowe (2000). We see
that the ab-type RR Lyraes in M33 (both the low reddening and higher
reddening samples)
are consistent with those in Oosterhoff I Galactic globular 
clusters. 
This suggests that the mean metallicities of these samples
are indistinguishable from each other. This reinforces the
assertion that the low reddening RR Lyraes as well as most of those 
that suffer from higher reddening are likely to be in the halo of M33.


Similar to our analysis in M31, it is possible to examine the radial metallicity gradient
in M33. Figure 7 is taken from the work of 
Yang et al. (2010) and shows a number of stellar populations that trace the disk
and halo of M33. Figure 7 shows the inner disk fields of Kim et al. (2002, solid circles) 
and the outer disk fields of Barker et al. (2007, open circles) both of which are based 
on red giant branch samples. The dotted line is the least-squares fit to the inner disk points from
Kim et al. (2002), which also reproduces the location of the outer disk points from 
Barker et al. (2007). The solid line in Figure 7 corresponds to the M33 planetary 
nebulae (PNe; Magrini et al. 2010), which are believed to be disk tracers with ages 
as old as 10 Gyr (Maraston 2005). Their abundances have been converted from 
12 + log (O/H) to [Fe/H] using the relation between [$\alpha$/Fe] and [Fe/H] from 
Barker \& Sarajedini (2008) for M33 and assuming that [$\alpha$/Fe] $\approx$ [O/Fe]. 
Also plotted in Figure 7 are representatives of the M33 halo population -- nine halo globular
clusters (open triangles) from Sarajedini et al. (2000) and the RRL variables from Sarajedini 
et al. (2006, open squares) and the present study (filled squares). The furthest M33 
globular cluster (EC-1) is also plotted (Stonkute et al. 2008). We note that this is a 
modified version of Figure 20 in the work of  Barker et al. (2007).

Figure 7 suggests that the metal abundance of the M33 disk decreases with 
increasing galactocentric distance. In contrast, the halo populations show no 
trend of metal abundance with radial position reminiscent of the halo globular 
cluster population in the Milky Way beyond 8 kpc from the Galactic center (Zinn 1985). 
This represents another piece of evidence supporting the assertion that the majority of the
M33 RR Lyraes are members of its halo rather than its disk.

\section{Summary and Conclusions}

Studies of RR Lyrae variables in M31 and M33
are still in their infancy but there are a few points we can make
with relative certainty. First, both galaxies do indeed harbor RR Lyrae
variables in their halos, which suggests that, similar to the Milky Way,
there is an old component
($\gea$10 Gyr) to the halos of M31 and M33. These variables have
pulsational properties that are consistent with those in Oosterhoff I
Galactic globular clusters. Furthermore, their mean metallicity
of the RR Lyraes in each galaxy is consistent with that of its halo
globular clusters. Much more work is left to be done.
For example, wide-field time-domain surveys of M31 and M33 will better
reveal the spatial distribution of their RR Lyraes. The capabilities of 30m 
telescopes will likely allow us to obtain kinematic and abundance information for
the RR Lyraes in M31 and M33. We can then study the detailed properties of the
earliest epochs of star formation in these galaxies. This 
has implications for how spiral galaxies fit into the $\Lambda$CDM
paradigm for structure formation in the early universe. 

\acknowledgements 
The author gratefully acknowledges the work of a number of close
collaborators that have contributed results to this review, especially Soung-Chul
Yang, whose doctoral dissertation at the University of Florida is the source of
much of the M33 RR Lyrae work. The template-fitting code
we use originated with Andy Layden and has been integrated into
our FITLC software by University of Florida
graduate student Conor Mancone. Much of this work has been
supported by NASA through
grants from the Space Telescope Science Institute which is operated by 
the Association of
Universities for Research in Astronomy, Incorporated, under NASA 
contract NAS5-26555.

\end{document}